\documentclass[aps,prl,
 amsmath,amssymb,
 groupedaddress,
 twocolumn,
 showpacs]{revtex4-1}
\usepackage{bm}
\usepackage{graphicx}
\usepackage{gensymb}
\usepackage{amsmath}
\usepackage{amssymb}
\usepackage{ulem}
\usepackage{here}

\begin{document}

\title{Fermiology of possible topological superconductor Tl$_{0.5}$Bi$_2$Te$_3$ derived from hole-doped topological insulator}
\author{C. X. Trang,$^1$ Z. Wang,$^2$ D. Takane,$^1$ K. Nakayama,$^1$ S. Souma,$^3$ T. Sato,$^1$ T. Takahashi,$^{1,3}$\\ A. A. Taskin,$^2$ and Yoichi Ando$^2$}

\affiliation{$^1$Department of Physics, Tohoku University, Sendai 980-8578, Japan\\
$^2$Institute of Physics II, University of Cologne, K$\ddot{o}$ln 50937, Germany\\
$^3$WPI Research Center, Advanced Institute for Materials Research, Tohoku University, Sendai 980-8577, Japan
}

\date{\today}

\begin{abstract}
    We have performed angle-resolved photoemission spectroscopy on Tl$_{0.5}$Bi$_2$Te$_3$, a possible topological superconductor derived from Bi$_2$Te$_3$. We found that the bulk Fermi surface consists of multiple three-dimensional hole pockets surrounding the {\it Z }point, produced by the direct hole doping into the valence band. The Dirac-cone surface state is well isolated from the bulk bands, and the surface chemical potential is variable in the entire band-gap range. Tl$_{0.5}$Bi$_2$Te$_3$ thus provides an excellent platform to realize two-dimensional topological superconductivity through a proximity effect from the superconducting bulk. Also, the observed Fermi-surface topology provides a concrete basis for constructing theoretical models for bulk topological superconductivity in hole-doped topological insulators.
\end{abstract}


\maketitle

Topological insulators (TIs) manifest a novel quantum state of matter in which a distinct topology of bulk wave functions produces a gapless surface or edge state across the bulk band gap inverted by a strong spin-orbit coupling. The discovery of TIs has stimulated intensive investigations into an even more exotic state of matter, a topological superconductor (TSC). TSCs generically possess a gapless Andreev bound state at the edge or surface which consists of Majorana fermions \cite{SchnyderPRB2008, SalommaPRB1988, ReadPRB2000, QiPRL2009, SatoPRB20092, SatoPRB2010, HasanReview, ZhangReview, AndoReview, AliceaRPP2012, Beenakker2013}. Owing to the peculiar characteristics of Majorana fermions that particle is its own antiparticle as well as its potential application to a fault-tolerant quantum computer, TSCs is one of the emergent topics in current condensed-matter physics.

While various schemes to realize TSCs have been proposed \cite{FuPRL2008, QiPRL2009, SatoPRB20092, SatoPRB2010, LinderPRL2010}, it has been suggested that a superconductor derived from a TI is a promising candidate for a TSC because of the strong spin-orbit coupling which would lead to unconventional electron pairing. The first of such a TSC candidate derived from a TI is Cu$_x$Bi$_2$Se$_3$ \cite{HorPRL2010}, in which the point-contact spectroscopy experiment detected a pronounced zero-bias conductance peak to signify unconventional surface Andreev bound states \cite{SasakiPRL2011}. Angle-resolved photoemission spectroscopy (ARPES) revealed that Cu-doping produces electron carriers in the inverted conduction band (CB) while leaving the Dirac-cone surface state (SS) isolated from the bulk band \cite{WrayNP2010}. Although these experimental observations would support either an unconventional odd-parity superconducting (SC) pairing relevant to the bulk topological superconductivity \cite{SatoPRB2009, SatoPRB2010, FuBergPRL2010} or the two-dimensional (2D) topological superconductivity {\it via} superconducting proximity effect \cite{FuPRL2008, WrayNP2010} the relationship between the electronic states and the topological superconductivity is still under intensive debates \cite{SasakiPRL2011, WrayNP2010, SatoPRB2009, FuBergPRL2010, KrienerPRL2011, KirzhnerPRB2012, TanakaPRB2012, LevyPRL2013, PengPRB2013, LahoudPRB2013, KrienerPRB2012, BayPRL2012}. Under this situation, it is of great importance to explore new platforms of TSCs by doping carriers into TIs in order to pin down the necessary conditions for realizing TSCs. However, only a limited number of candidate materials, such as Cu$_x$(PbSe)$_5$(Bi$_2$Se$_3$)$_6$ \cite{SasakiPRBR2014, NakayamaPRBR2015}, Sr$_x$Bi$_2$Se$_3$ \cite{LiuJACS2015}, and Nb$_x$Bi$_2$Se$_3$ \cite{AsabaArxiv} have been discovered in this materials class till now. It is useful to note that a hole-doped topological crystalline insulator, In$_x$Sn$_{1-x}$Te, is also found to be a promising candidate of a TSC  \cite{SasakiPRL2012, NovakPRB2013, SatoPRL2013}.
 
 Recently, it was discovered that a relatively large amount of thallium (Tl) doping into a prototypical TI, Bi$_2$Te$_3$, gives rise to the superconductivity of $T_c$ = 2.28 K with the SC volume fraction as high as 95$\%$ \cite{WangCM2016}. Tl$_x$Bi$_2$Te$_3$ shows some intriguing characteristics distinct from Cu$_x$Bi$_2$Se$_3$. For example, the Tl ions are not ``intercalated" but ``incorporated" into the tetradymite matrix, in contrast to the case of Cu$_x$Bi$_2$Se$_3$ in which Cu ions are intercalated between two Bi$_2$Se$_3$ quintuple layers. The Hall-resistivity measurements signified the dominance of $p$-type carriers in TI$_x$Bi$_2$Te$_3$, whereas other TI-based superconductors \cite{HorPRL2010, SasakiPRBR2014, LiuJACS2015} are $n$-type. Thus, it is of significant importance to experimentally establish the electronic structure of Tl$_x$Bi$_2$Te$_3$ to investigate the possible topological superconductivity.
 
 In this paper, we report high-resolution ARPES on Tl$_x$Bi$_2$Te$_3$ ($x$ = 0.5). By utilizing energy-tunable photons from synchrotron radiation, we determined the bulk and surface electronic states separately. We found that the bulk electronic states are heavily hole-doped, as evident from the existence of multiple three-dimensional (3D) hole pockets. Also, the 2D topological SS was found to be well isolated from the bulk bands. We discuss the implications of the present results to the possible topological superconductivity in Tl$_x$Bi$_2$Te$_3$ by comparing with Cu$_x$Bi$_2$Se$_3$.

 High-quality single crystals of Tl$_x$Bi$_2$Te$_3$ were synthesized from high-purity elemental shots of Tl (99.99$\%$), Bi (99.9999$\%$) and Te (99.9999$\%$). Details of the sample preparation were described elsewhere \cite{WangCM2016}.  We performed ARPES measurements on two samples with $x$ = 0.5 and 0.6. While the obtained ARPES data show no intrinsic differences between the two, the data quality appears to be slightly better for the $x$ = 0.5 sample used for this study (see Supplemental Material for details), which is likely due to the different fraction of impurity phases in crystals \cite{WangCM2016}.  It is noted that the SC volume fraction for $x$ = 0.5 is 64$\%$ \cite{WangCM2016}, reasonably high for discussing the electronic states relevant to the superconductivity. ARPES measurements were performed with MBS-A1 and Omicron-Scienta SES2002 electron analyzers with energy-tunable synchrotron lights at BL-7U in UVSOR and BL28A in Photon Factory. We used the linearly polarized light of 15--30 eV and the circularly polarized one of 40--90 eV. The energy and angular resolutions were set at 10--30 meV and 0.2$^\circ$, respectively. Samples were cleaved {\it in situ} along the (111) crystal plane [see Fig. 1(a)] in an ultrahigh vacuum of 1$\times$10$^{-10}$ Torr, and kept at $T$ = 40 K during the ARPES measurements. A shiny mirror-like surface was obtained after cleaving to show the high quality of samples. The Fermi level ($E_{\rm F}$) of samples was referenced to that of a gold film evaporated onto the sample holder. The first-principles band-structure calculations including the spin-orbit interaction were carried out on bulk Bi$_2$Te$_3$ by using the Quantum-Espresso code \cite{Giannozzi}. The plane-wave cutoff energy and the $k$-point mesh were set to be 30 Ry and 8 $\times$ 8 $\times$ 8, respectively.
   
  \begin{figure}
 \includegraphics[width=3in]{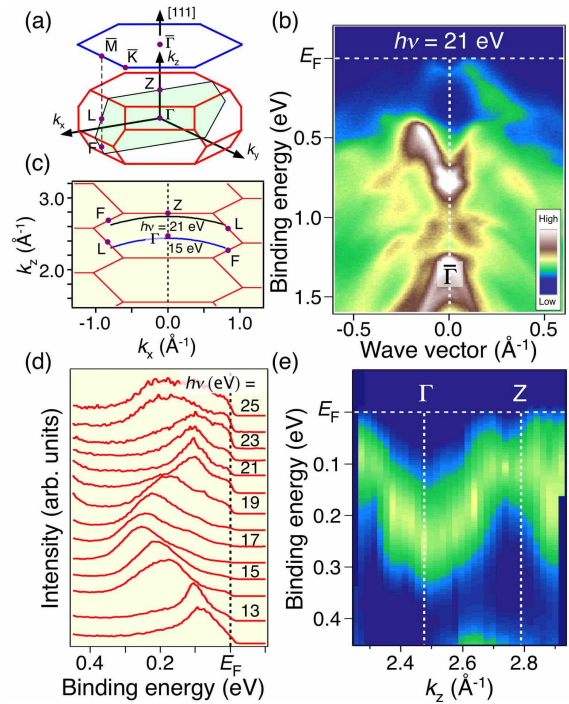}
\caption{(color online). (a) Bulk rhombohedral BZ (red) and corresponding hexagonal surface BZ (blue) of Tl$_{0.5}$Bi$_2$Te$_3$. Green shaded area indicates the $k_{x}$-$k_{z}$ plane. (b) Plot of ARPES intensity in the VB region as a function of the in-plane wave vector and $E_{\rm B}$, measured at $h\nu$ = 21 eV at $T$ = 40 K. (c) Bulk BZ in the $k_{x}$-$k_{z}$ plane together with measured cuts for $h\nu$ = 15 and 21 eV (blue and black curves). $k_{z}$ values were calculated with the inner potential $V_{\rm 0}$ of 12 eV which was estimated from the periodicity of bands in (d). (d) Photon-energy dependence of the normal-emission EDCs near $E_{\rm F}$. (e) Plot of ARPES intensity as a function of $k_{z}$ and $E_{\rm B}$. The location of $\Gamma$ and {\it Z} points in bulk BZ is indicated by dashed lines.}
\end{figure}

 First, we discuss the overall valence-band (VB) structure of Tl$_{0.5}$Bi$_2$Te$_3$. Figure 1(b) displays the plot of ARPES intensity around the $\bar{\Gamma}$ point as a function of in-plane wave vector and binding energy ($E_{\rm B}$), measured with photons of $h\nu$ = 21 eV. One can see several intensive features at $E_{\rm B}$ = 0.4--1.5 eV due to the Te 5$p$ states \cite{ZhangNP2009} and a weaker holelike band near $E_{\rm F}$. To determine the band dispersions of the VB in the 3D $k$-space, we have performed $h\nu$-dependent ARPES measurements. Figure 1(d) shows the energy distribution curves (EDCs) at the normal emission measured with various photon energies from 12 to 25 eV. The topmost VB located at $E_{\rm B}${$\sim$}0.1 eV at $h\nu$ = 21--24 eV disperses toward higher $E_{\rm B}$'s on decreasing $h\nu$, stays at $E_{\rm B}${$\sim$}0.25 eV at $h\nu$ = 15-17 eV, and disperses back again toward $E_{\rm F}$ on further decreasing $h\nu$. Such a periodic nature of the band dispersion is more clearly visualized in the ARPES-intensity plot along the wave vector perpendicular to the sample surface ($k_{z}$), as shown in Fig. 1(e). Obviously, the highest occupied VB has a bottom at the $\Gamma$ point of the bulk Brillouin zone (BZ), while its top is located around the {\it Z} point.

\begin{figure*}[htbp]
 \includegraphics[width=6.8in]{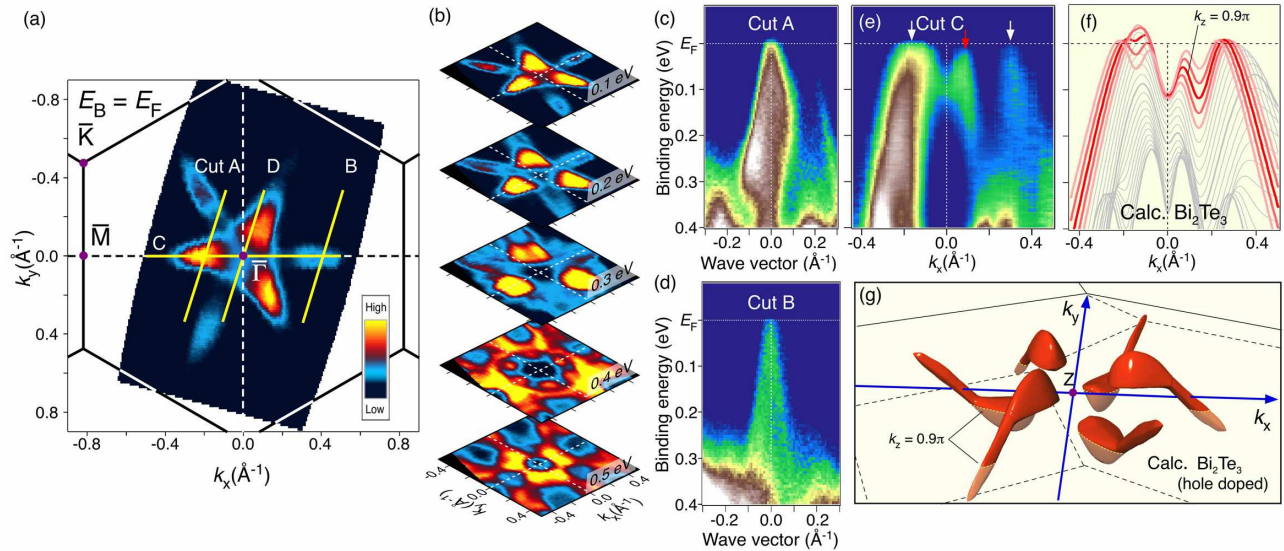}
\caption{(color online). (a) ARPES-intensity mapping at $E_{\rm F}$ as a function of the 2D wave vector ($k_{x}$-$k_{y}$) measured at $h\nu$ = 21 eV. Yellow lines indicate the measured \textbf{k} cuts. (b) ARPES-intensity mapping for various $E_{\rm B}$'s between 0.1 and 0.5 eV. (c)-(e) Near-$E_{\rm F}$ ARPES intensity as a function of the in-plane wave vector and $E_{\rm B}$ measured along representative \textbf{k} cuts (A-C) shown in (a). (f) Calculated band structure along the $\bar{\Gamma}\bar{M}$ cut obtained from the first-principles band calculations of bulk Bi$_2$Te$_3$ for various $k_{z}$'s. In order to take into account the hole-doping effect, the energy position of $E_{\rm F}$ was set to be 70 meV below the VB top in the calculation. Band dispersions centered at $k_{z}$ = 0.9 $\pi$ are shown by red curves; gradual shading highlights the $k_{z}$ broadening of the $\pm$ 0.1 $\pi$ window. (g) 3D view of a schematic FS around the {\it Z} point, constructed from the band calculations of hole-doped Bi$_2$Te$_3$ with the $E_{\rm F}$ position at 70 meV below the VB top. Yellow and red colors on the FS corresponds to portions for $k_{z}$ $<$ 0.9 $\pi$ and $k_{z}$ $>$ 0.9 $\pi$, respectively.}
\end{figure*}

To see more clearly the electronic states responsible for the formation of the Fermi surface (FS), we have chosen $h\nu$ = 21 eV whose $\bf{k}$ cut nearly crosses the {\it Z} point [see Fig. 1(c)], and mapped out the FS as a function of the 2D wave vector ($k_{x}$ and $k_{y}$). As seen in Fig. 2(a), the FS consists of three large and three small ellipsoidal pockets elongated in the $\bar{\Gamma}\bar{M}$ direction. It is noted that the arrangement and the intensity of FSs exhibit a three-fold pattern obeying the bulk-crystal symmetry. Both the large and small FSs commonly have a holelike character, as recognized from the gradual expansion of the ellipsoid in the equi-energy cut on increasing $E_{\rm B}$ as shown in Fig. 2(b). In fact, the band dispersion along $\bf{k}$ cuts crossing these pockets [cuts A and B in Fig. 2(a)] signifies a $\Lambda$-shaped intensity pattern, consistent with the holelike nature [see Figs. 2(c) and 2(d)]. Figure 2(e) shows the ARPES-derived band dispersions along the $\bar{\Gamma}\bar{M}$ cut [cut C in Fig. 2(a)].  We find a reasonable agreement between the experimental band dispersions and the first-principles band-structure calculations [Fig. 2(f)]. In particular, the $\bf{k}$ location of the large and small hole pockets (white arrows) as well as a small hump at $k_{x}$ = 0.07 \AA$^{-1}$ (red arrow) in the experiment are well reproduced in the calculation when the chemical potential is located at 70 meV below the VB top. This demonstrates that the emergence of multiple hole pockets is a consequence of a direct hole doping into the VB in Bi$_2$Te$_3$. Based on this good agreement between the experiment and the calculation, we have constructed the 3D bulk FSs in Tl$_{0.5}$Bi$_2$Te$_3$ by calculating a series of Fermi wave vectors ($k_{\rm F}$'s) for hole-doped Bi$_2$Te$_3$, as shown in Fig. 2(g). One can clearly see multiple FSs surrounding the {\it Z} point, which consists of six equivalent tadpole-shaped hole pockets reflecting the bulk-BZ symmetry. The ``body" of tadpole FS is located near the {\it Z} point and its ``tail" extends away from the {\it Z} point. When we consider a fixed $k_{z}$ plane slightly below the {\it Z} point ($k_{z}$ = 0.9 $\pi$) to account for the experimental $k_{z}$ value for $h\nu$ = 21 eV, this plane slices three ``bodies" and three ``tails" of the six tadpole pockets. Intriguingly, such features are clearly observed in the experimental FS mapping in Fig. 2(a), in which one can attribute three pockets with strong (weak) intensity close to (away from) the {\it Z} point to the body (tail) of the tadpole FS.

\begin{figure}[htbp]
 \includegraphics[width=3.4in]{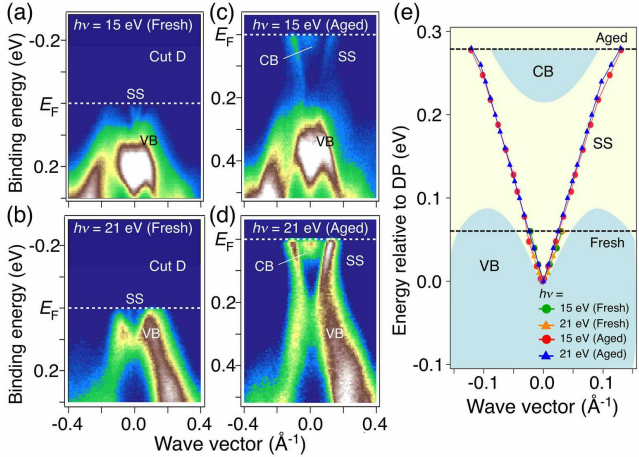}
 \caption{(color online). (a), (b) Near-$E_{\rm F}$ ARPES intensity just after cleaving (fresh), measured along cut D in Fig. 2(a) at $h\nu$ = 15 and 21 eV, respectively. $\bf{k}$ cuts at $h\nu$ = 15 and 21 eV nearly cross the $\Gamma$ and {\it Z} points of bulk BZ, respectively, as shown in Fig. 1(c). VB and SS denote the valence band and surface state, respectively. (c), (d) The same as (a) and (b) but 12 hours after cleaving (aged). CB denotes the conduction band. (e) Dirac-cone band dispersion relative to the Dirac point (DP) for fresh and aged surfaces of Tl$_{0.5}$Bi$_2$Te$_3$. The energy position of the chemical potential $\mu$ for each sample is indicated by a horizontal dashed line. The band dispersion was estimated from the numerical fittings of the momentum distribution curves for (a)-(d).}
\end{figure}

Having established the bulk FS topology, the next important question is whether or not the topological SS exists in Tl$_{0.5}$Bi$_2$Te$_3$. Figure 3(a) shows the plot of ARPES intensity along a cut crossing the $\bar{\Gamma}$ point [cut D in Fig. 2(a)] measured at $h\nu$ = 15 eV [see Fig. 1(c)]. Besides the M-shaped VB feature, one can recognize a V-shaped electronlike band in the vicinity of $E_{\rm F}$ inside the valley of the VB. This tiny feature is assigned to a SS since its energy location is stationary when the photon energy is changed to $h\nu$ = 21 eV as seen in Fig. 3(b). This indicates that the band-inverted nature is conserved even upon the Tl doping. Figure 3(e) shows directly that the band dispersions obtained with these two different photon energies, $h\nu$ = 15 and 21 eV, exactly overlap with each other [Fig. 3(e)]. Furthermore, as shown in Figs. 3(c) and 3(d), we were able to trace the band dispersion of the SS up to $\sim$ 0.2 eV above the VB top [see also Fig. 3(e)] simply by keeping the samples in a vacuum of 2$\times$10$^{-10}$ Torr for 12 hours, which resulted in a significant downward surface band-bending. This indicates that electron doping due to surface aging can easily shift the surface chemical potential up to the bulk CB.

The above results imply that the Tl$_x$Bi$_2$Te$_3$ system provides an excellent platform to realize the 2D topological superconductivity. Since the surface chemical potential can be systematically controlled by the surface-aging technique throughout the entire band-gap range as demonstrated in Fig. 3, it is possible to isolate the topological SS from the VB and CB in both the $k$ space and the energy axis by controlling the strength of the surface band-bending. This is in favor of the realization of 2D topological superconductivity in the topological SS through the proximity effect from the SC bulk, which requires the SS to be well separated from the bulk bands \cite{WrayNP2010}. In this regard, the electronic structure as well as the mobile nature of the surface chemical potential in Tl$_x$Bi$_2$Te$_3$ is more desirable than those of Cu$_x$Bi$_2$Se$_3$, in which the surface chemical potential is pinned deep inside the CB and is hardly controlled by the surface aging.
 
 Besides the prospect for realizing 2D topological superconductivity on the surface, it is possible that the bulk superconductivity in this material is topological. In this regard, the most intriguing aspect of Tl$_x$Bi$_2$Te$_3$ is its hole-doped nature, which is in contrast to the electron-doped nature of the other TI-derived superconductors discovered so far, as highlighted in Fig. 4 by a comparison of the electronic states in Tl$_x$Bi$_2$Te$_3$ and Cu$_x$Bi$_2$Se$_3$. It is known that a sufficient condition to realize 3D TSC is (i) the odd-parity SC pairing and (ii) the FS in the normal state enclosing an odd number of time-reversal invariant momenta (TRIM) \cite{SatoPRB2010, FuBergPRL2010}. The situation (ii) is realized in lightly electron-doped Cu$_x$Bi$_2$Se$_3$ where the bulk spherical electron pocket encloses a TRIM, the $\Gamma$ point \cite{WrayNP2010, TanakaPRB2012, LahoudPRB2013}, whereas the FSs in Tl$_x$Bi$_2$Te$_3$ enclose no TRIM [compare the top panels of Figs. 4(a) and 4(b)]. As for the requirement (i), the structure of the effective Dirac Hamiltonian for Bi$_2$Se$_3$ was shown to allow SC pairing between two different $p$ orbitals possessing opposite parity, leading to an odd-parity pairing \cite{FuBergPRL2010}. Since the effective Hamiltonian is the same for Bi$_2$Te$_3$ \cite{ZhangNP2009, ZhangPRB2010}, the same argument for odd-parity pairing would apply to Tl$_x$Bi$_2$Te$_3$. However, since the FSs of Tl$_x$Bi$_2$Te$_3$ do not enclose any TRIMs, judging the possibility of topological superconductivity requires heuristic efforts to find a concrete topological invariant. In this regard, our observation of the FS topology in Tl$_{0.5}$Bi$_2$Te$_3$ provides a solid experimental basis for constructing theoretical models for possible 3D topological superconductivity in Tl$_x$Bi$_2$Te$_3$ and more generally in a hole-doped counterpart of TIs.
 
\begin{figure}[tbp]
\includegraphics[width=3.4in]{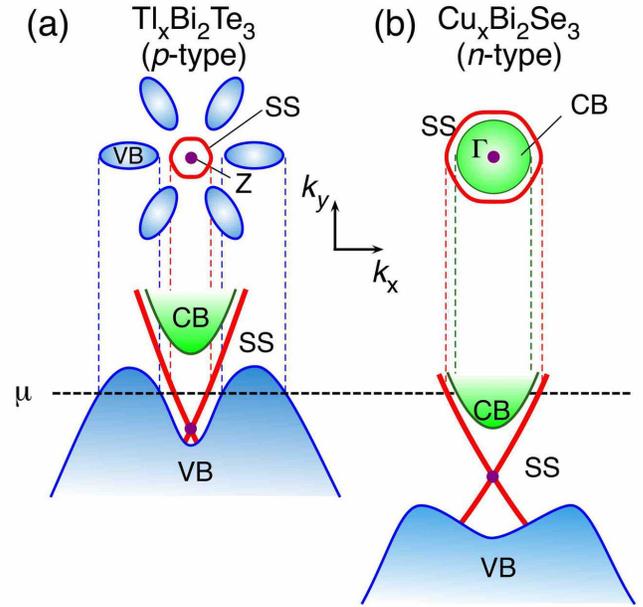}
\caption{(color online). Comparison of the schematic FS (top) and the band diagram (bottom) between (a) Tl$_x$Bi$_2$Te$_3$ and (b) Cu$_x$Bi$_2$Se$_3$.}
 \end{figure}

To summarize, we report high-resolution ARPES results on Tl$_{0.5}$Bi$_2$Te$_3$ to elucidate the energy-band structure underlying the SC states. By utilizing energy-tunable photons, we have succeeded in experimentally determining the electronic structure of both bulk and surface states. The bulk FS consists of a set of hole pockets, establishing that Tl$_{0.5}$Bi$_2$Te$_3$ is a rare case of a superconductor derived by hole doping into TIs. We also found that the topological SS is well isolated from the bulk bands and the surface chemical potential is easily tunable over the entire band-gap region, which makes this material a suitable candidate to realize 2D topological superconductivity on the surface. Also, our determination of the bulk FS topology lays a foundation for constructing theoretical models for 3D topological superconductivity in hole-doped tetradymite TIs.

\begin{acknowledgements}
We thank Y. Tanaka, H. Kimizuka, N. Inami, H. Kumigashira, K. Ono, S. Ideta, and K. Tanaka for their assistance in the ARPES measurements and K. Akagi for his assistance in the band calculations. This work was supported by MEXT of Japan (Innovative Area ``Topological Materials Science", 15H05853), JSPS (KAKENHI 15H02105, 26287071, 25287079), KEK-PF (Proposal No. 2015S2-003), and UVSOR (Proposal No. 27-807).
\end{acknowledgements}


\bibliographystyle{prsty}

\begin{thebibliography}{30}

\bibitem{SchnyderPRB2008} A. P. Schnyder, S. Ryu, A. Furusaki, and A.W.W. Ludwig, Phys. Rev. B  \textbf{78}, 195125 (2008).

\bibitem{SalommaPRB1988} M. M. Salomaa and G. E. Volovik, Phys. Rev. B  \textbf{37}, 9298 (1988).

\bibitem{ReadPRB2000} N. Read and D. Green, Phys. Rev. B  \textbf{61}, 10267 (2000).

\bibitem{QiPRL2009}X.-L. Qi, T. L. Hughes, S. Raghu, and S.-C. Zhang, Phys. Rev. Lett.  \textbf{102}, 187001 (2009).

\bibitem{SatoPRB20092} M. Sato and S. Fujimoto, Phys. Rev. B  \textbf{79}, 094504 (2009).

\bibitem{SatoPRB2010} M. Sato, Phys. Rev. B  \textbf{81}, 220504(R) (2010).

\bibitem{HasanReview} M. Z. Hasan and C. L. Kane, Rev. Mod. Phys.  \textbf{82}, 3045 (2010).

\bibitem{ZhangReview} X.-L. Qi and S.-C. Zhang, Rev. Mod. Phys.  \textbf{83}, 1057 (2011).

\bibitem{AndoReview} Y. Ando, J. Phys. Soc. Jpn.  \textbf{82}, 102001 (2013).

\bibitem{AliceaRPP2012} J. Alicea, Rep. Prog. Phys. \textbf{75}, 076501 (2012).

\bibitem{Beenakker2013} C. W. J. Beenakker, Annu. Rev. Condens. Matter Phys. \textbf{4}, 113 (2013).

\bibitem{FuPRL2008} L. Fu and C. L. Kane, Phys. Rev. Lett. \textbf{100}, 096407 (2008).

\bibitem{LinderPRL2010} J. Linder, Y. Tanaka, T. Yokoyama, A. Sudbo, and N. Nagaosa, Phys. Rev. Lett. \textbf{104}, 067001 (2010).

\bibitem{HorPRL2010} Y. S. Hor, A. J. Williams, J. G. Checkelsky, P. Roushan, J. Seo, Q. Xu, H. W. Zandbergen, A. Yazdani, N. P. Ong, and R. J. Cava, Phys. Rev. Lett. \textbf{104}, 057001 (2010).

\bibitem{SasakiPRL2011} S. Sasaki, M. Kriener, K. Segawa, K. Yada, Y. Tanaka, M. Sato, and Y. Ando, Phys. Rev. Lett. \textbf{107}, 217001 (2011).

\bibitem{WrayNP2010} L. A. Wray, S.-Y. Xu, Y. Xia, Y. S. Hor, D. Qian, A. V. Fedorov, H. Lin, A. Bansil, R. J. Cava, and M. Z. Hasan, Nat. Phys. \textbf{6}, 855 (2010).

\bibitem{SatoPRB2009} M. Sato, Phys. Rev. B \textbf{7}9, 214526 (2009).

\bibitem{FuBergPRL2010} L. Fu and E. Berg, Phys. Rev. Lett. \textbf{105}, 097001 (2010).

\bibitem{KrienerPRL2011} M. Kriener, K. Segawa, Z. Ren, S. Sasaki, and Y. Ando, Phys. Rev. Lett. \textbf{106}, 127004 (2011).


\bibitem{KirzhnerPRB2012} T. Kirzhner, E. Lahoud, K. B. Chaska, Z. Salman, and A. Kanigel, Phys. Rev. B \textbf{86}, 064517 (2012).

\bibitem{TanakaPRB2012} Y. Tanaka, K. Nakayama, S. Souma, T. Sato, N. Xu, P. Zhang, P. Richard, H. Ding, Y. Suzuki, P. Das, K. Kadowaki, and T. Takahashi, Phys Rev. B \textbf{85}, 125111 (2012).

\bibitem{LevyPRL2013} N. Levy, T. Zhang, J. Ha, F. Sharifi, A. A. Talin, Y. Kuk, and J. A. Stroscio, Phys. Rev. Lett. \textbf{110}, 117001 (2013).

\bibitem{PengPRB2013} H. Peng, D. De, B. Lv, F. Wei, and C.-W. Chu, Phys. Rev. B \textbf{88}, 024515 (2013).

\bibitem{LahoudPRB2013} E. Lahoud, E. Maniv, M. S. Petrushevsky, M. Naamneh, A. Ribak, S. Wiedmann, L. Petaccia, Z. Salman, K. B. Chashka, Y. Dagan, and A. Kanigel, Phys. Rev. B \textbf{88}, 195107 (2013).

\bibitem{KrienerPRB2012} M. Kriener, K. Segawa, S. Sasaki, and Y. Ando, Phys. Rev. B \textbf{86}, 180505(R) (2012).

\bibitem{BayPRL2012} T. V. Bay, T. Naka, Y. K. Huang, H. Luigjes, M. S. Golden, and A. de Visser, Phys. Rev. Lett. \textbf{108}, 057001 (2012).

\bibitem{SasakiPRBR2014} S. Sasaki, K. Segawa, and Y. Ando, Phys. Rev. B \textbf{90}, 220504(R) (2014).

\bibitem{NakayamaPRBR2015} K. Nakayama, H. Kimizuka, Y. Tanaka, T. Sato, S. Souma, T. Takahashi, S. Sasaki, K. Segawa, and Y. Ando, Phys. Rev. B \textbf{92}, 100508(R) (2015).

\bibitem{LiuJACS2015} Z. Liu, X. Yao, J. Shao, M. Zuo, L. Pi, S. Tan, C. Zhang, and Y. Zhang, J. Am. Chem. Soc. \textbf{137}, 10512 (2015).

\bibitem{AsabaArxiv} T. Asaba, B. J. Lawson, C. Tinsman, L. Chen, P. Corbae, G. Li, Y. Qiu, Y. S. Hor, L. Fu, and L. Li, arXiv:1603.04040 (2016).

\bibitem{SasakiPRL2012} S. Sasaki, Z. Ren, A. A. Taskin, K. Segawa, L. Fu, and Y. Ando, Phys. Rev. Lett. \textbf{109}, 217004 (2012).

\bibitem{NovakPRB2013} M. Novak, S. Sasaki, M. Kriener, K. Segawa, and Y. Ando, Phys. Rev. B \textbf{88}, 140502(R) (2013).

\bibitem{SatoPRL2013} T. Sato, Y. Tanaka, K. Nakayama, S. Souma, T. Takahashi, S. Sasaki, Z. Ren, A. A. Taskin, K. Segawa, and Y. Ando, Phys. Rev. Lett. \textbf{110}, 206804 (2013).

\bibitem{WangCM2016} Z. Wang, A. A. Taskin, T. Fr\"olich, M. Braden, and Y. Ando, Chem. Mater. \textbf{28}, 779 (2016).

\bibitem{Giannozzi} P. Giannozzi {\it et al.}, J. Phys.: Condens. Matter. \textbf{21}, 395502 (2009).

\bibitem{ZhangNP2009} H. Zhang, C.-X. Liu, X.-L. Qi, X. Dai, Z. Fang, and S.-C. Zhang, Nat. Phys. \textbf{5}, 438 (2009).

\bibitem{ZhangPRB2010} C.-X. Liu, X.-L. Qi, H. J. Zhang, X. Dai, Z. Fang, and S.-C. Zhang, Phys. Rev. B \textbf{82}, 045122 (2010).


 
\end{thebibliography}

\clearpage

{

\onecolumngrid

\begin{center}
{\large Supplemental Materials for \\
``Fermiology of possible topological superconductor Tl$_{0.5}$Bi$_2$Te$_3$ \\derived from hole-doped topological insulator''}

\vspace{0.3 cm}

C. X. Trang,$^1$ Z. Wang,$^2$ D. Takane,$^1$ K. Nakayama,$^1$ S. Souma,$^3$ T. Sato,$^1$ T. Takahashi,$^{1,3}$\\ A. A. Taskin,$^2$ and Yoichi Ando$^2$

{\footnotesize
$^1${\it Department of Physics, Tohoku University, Sendai 980-8578, Japan}
\newline
$^2${\it Institute of Physics II, University of Cologne, K$\ddot{o}$ln 50937, Germany}
\newline
$^3${\it WPI Research Center, Advanced Institute for Materials Research, Tohoku University, Sendai 980-8577, Japan}
}

\end{center}


\renewcommand{\thefigure}{S\arabic{figure}}
\setcounter{figure}{0}

\subsection{S1. Comparison of band structure between x = 0.5 and 0.6}
Figure S1 displays a comparison of the ARPES intensity near the Fermi level ($E_{\rm F}$) measured along a $k$ cut crossing the $\bar{\Gamma}$ point of the surface Brillouin zone at $h\nu$ = 15 eV. At $x$ = 0.5, one can clearly resolve a signature of a V-shaped surface state (SS) inside the M-shaped bulk valence band (VB). As seen from a side-by-side comparison of the intensity pattern, the overall spectral feature is essentially the same between $x$ = 0.5 and 0.6, whereas the spectral intensity is slightly broader in $x$ = 0.6. It has been reported from the powder x-ray diffraction experiments of Tl$_x$Bi$_2$Te$_3$ ($x$ = 0.0-0.6) that the volume of impurity phases (which mainly arise from TlBiTe$_2$ phase) gradually increases with increasing $x$ [34]. Thus, the broader spectral feature for $x$ = 0.6 likely originates from the shorter lifetime of surface quasiparticles due to stronger scattering by impurities.

\begin{figure}[H]
\begin{center}
\includegraphics[width=3.4 in]{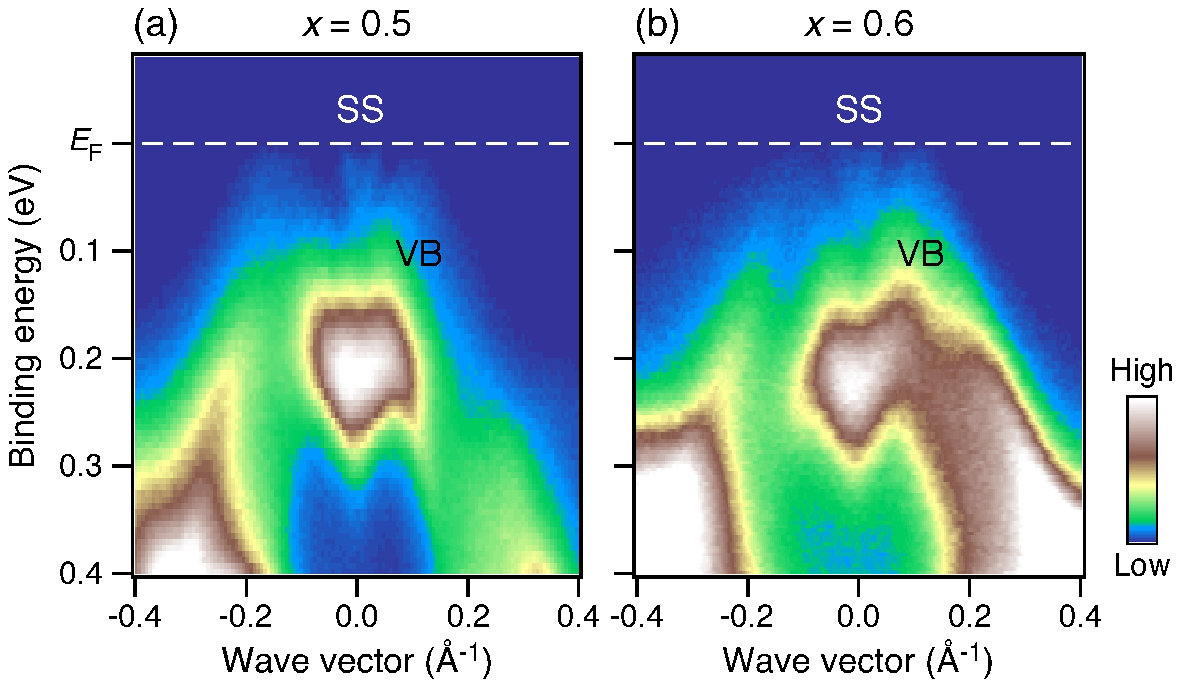}
\caption{(a), (b) Plot of near-$E_{\rm F}$ ARPES intensity in a $k$ cut crossing $\bar{\Gamma}$ point as a function of the in-plane wave vector and binding energy for Tl$_x$Bi$_2$Te$_3$ with $x$ = 0.5 and 0.6, respectively. The data were recorded at $h\nu$ = 15 eV at $T$ = 40 K.
}
\end{center}
\end{figure}

\subsection {REFERENCES}
{
\noindent
[34] Z. Wang, A. A. Taskin, T. Fr$\ddot{o}$lich, M. Braden, and Y. Ando, Chem. Mater.  \textbf{28}, 779 (2016).
}

}

\end{document}